\documentclass[11pt]{article}
\usepackage{epsfig}
\def\nn{\nonumber}
\def\be{\begin{equation}}
\def\ee{\end{equation}}
\newcommand{\bea}{\begin{eqnarray}}
\newcommand{\eea}{\end{eqnarray}}
\newcommand{\bdm}{\begin{displaymath}}
\newcommand{\edm}{\end{displaymath}}
\long\def\symbolfootnote[#1]#2{\begingroup%
\def\thefootnote{\fnsymbol{footnote}}\footnote[#1]{#2}\endgroup}

\def\sdbeta{s_{2\beta}}

\def\sq2{\sqrt{2}}
\def\drbar{\overline{\rm DR}}
\def\msbar{\overline{\rm MS}}
\def\smalldrbar{\scriptscriptstyle{\overline{\rm DR}}}

\def\tb{\tan\beta}
\def\cb{c_\beta}
\def\sb{s_\beta}
\def\gl{\tilde{g}}
\def\mg{m_{\gl}}

\def\hsm{H_{\scriptscriptstyle{\rm SM}}}
\def\gam{Z}
\def\x1g{x_{1}}

\newcommand{\as}{\alpha_s}
\newcommand{\smallz}{{\scriptscriptstyle Z}} 
\newcommand{\smallw}{{\scriptscriptstyle W}} %
\newcommand{\smallh}{{\scriptscriptstyle H}} %
\newcommand{\smallr}{{\scriptscriptstyle R}} %
\newcommand{\mz}{m_\smallz}

\newcommand{\mh}{m_\smallh}
\newcommand{\muF}{\mu_{\scriptscriptstyle F}}
\newcommand{\muR}{\mu_\smallr}

\def\mt{m_t}

\def\sdt{s_{2\theta_t}}

\def\mb{m_b}
\def\sbu{\tilde{b}_1}
\def\sbd{\tilde{b}_2}

\def\bu{m_{\tilde{b}_1}^2}
\def\bd{m_{\tilde{b}_2}^2}

\def\buq{m_{\tilde{b}_1}^4}
\def\bdq{m_{\tilde{b}_2}^4}

\def\sdb{s_{2\theta_b}}
\def\cdb{c_{2\theta_b}}

\newenvironment{appendletterA}
 {
  \setcounter{section}{0}
  \setcounter{equation}{0}
  
 }{
 }
\newenvironment{appendletterB}
 {
  \setcounter{equation}{0}
  
 }{
 }

\begin{document}

\begin{titlepage}


{\flushright{
        \begin{minipage}{5cm}
          RM3-TH/10-16
        \end{minipage}        }

}
\renewcommand{\thefootnote}{\fnsymbol{footnote}}
\vskip 2cm
\begin{center}
\boldmath
{\LARGE\bf NLO QCD bottom corrections \\[7pt]
to Higgs boson production in the MSSM}\unboldmath
\vskip 1.cm
{\Large{G.~Degrassi$^{a}$ and P.~Slavich$^{b}$}}
\vspace*{8mm} \\
{\sl ${}^a$
    Dipartimento di Fisica, Universit\`a di Roma Tre and  INFN, Sezione di
    Roma Tre \\
    Via della Vasca Navale~84, I-00146 Rome, Italy}
\vspace*{2.5mm}\\
{\sl ${}^b$  LPTHE, 4, Place Jussieu, F-75252 Paris,  France}
\end{center}
\symbolfootnote[0]{{\tt e-mail:}}
\symbolfootnote[0]{{\tt degrassi@fis.uniroma3.it}}
\symbolfootnote[0]{{\tt slavich@lpthe.jussieu.fr}}

\vskip 0.7cm

\begin{abstract}
  We present a calculation of the two-loop bottom-sbottom-gluino
  contributions to Higgs boson production via gluon fusion in the
  MSSM.  The calculation is based on an asymptotic expansion in the
  masses of the supersymmetric particles, which are assumed to be much
  heavier than the bottom quark and the Higgs bosons.  We obtain
  explicit analytic results that allow for a straightforward
  identification of the dominant contributions in the NLO bottom
  corrections. We emphasize the interplay between the calculations of
  the masses and the production cross sections of the Higgs bosons,
  discussing sensible choices of renormalization scheme for the
  parameters in the bottom/sbottom sector.
\end{abstract}
\vfill
\end{titlepage}    
\setcounter{footnote}{0}

\section{Introduction}

With the coming into operation of the Large Hadron Collider (LHC), a
new era has begun in the search for the Higgs boson(s). At the LHC the
main production mechanism for the Standard Model (SM) Higgs boson,
$\hsm$, is the loop-induced gluon fusion mechanism \cite{H2gQCD0}, $gg
\to \hsm$, where the coupling of the gluons to the Higgs is mediated
by loops of colored fermions, primarily the top quark. The knowledge
of this process in the SM includes the full next-to-leading order
(NLO) QCD corrections \cite{H2gQCD1,SDGZ}, the next-to-next-to-leading
order (NNLO) QCD corrections \cite{H2gQCD2} including finite top mass
effects \cite{H2gQCD3}, soft-gluon resummation effects \cite{bd4}, an
estimate of the next-to-next-to-next-to-leading order (NNNLO) QCD
effects \cite{Moch:2005ky} and also the first-order electroweak
corrections \cite{DjG,ABDV0,APSU}.

The Minimal Supersymmetric extension of the Standard Model, or MSSM,
features a richer Higgs spectrum which consists of two neutral CP-even
bosons $h, H$, one neutral CP-odd boson $A$ and two charged scalars
$H^\pm$. The gluon-fusion process is one of the most important
production mechanisms for the neutral Higgs bosons, whose couplings to
the gluons are mediated by colored fermions and their supersymmetric
partners. The gluon-fusion cross section in the MSSM is known at the
NLO. The contributions arising from diagrams with squarks and gluons
were first computed under the assumption of vanishing Higgs mass in
ref.~\cite{Dawson:1996xz}. The complete top/stop contributions,
including stop mixing and gluino effects, were computed under the same
assumption in ref.~\cite{HS}, and the result was cast in a compact
analytic form in ref.~\cite{DS}. Later, more refined calculations
aimed at the inclusion of the full Higgs-mass dependence. In
particular, the full squark-gluon contribution is known in a closed
analytic form \cite{babis1,ABDV,MS}, while the full
quark-squark-gluino contribution has been computed in
ref.~\cite{babis2} via a combination of analytic and numerical
methods.

It should be stressed that, at least for the case of the light Higgs,
the exact two-loop QCD Higgs-gluon-gluon amplitude is in general well
approximated by the amplitude evaluated in the limit of neglecting the
Higgs mass. The latter is much easier to compute and the corresponding
result can be straightforwardly implemented in computer codes that aim
to evaluate the Higgs boson production cross section in a fast and efficient
way. Indeed, it was noticed several years ago for the SM case
\cite{KLS} that the exact $K$ factor, defined as the ratio between the
NLO and leading-order (LO) cross sections, is well approximated by the
so-called effective $K$ factor that can be obtained via an improved
effective-theory calculation. By the latter we mean a result in which
the effective NLO cross section is obtained by multiplying the exact
LO partonic cross section by the ${\cal O}(\as)$ corrections evaluated
in the  limit of vanishing Higgs mass.  For the SM case this approximation
works at the level of few per cent for Higgs mass values below the
$2\, \mt$ threshold, and up to $10\%$ for any Higgs mass value. The
same level of accuracy is reached when the Higgs couples to a generic
scalar particle with mass $m_S$, with the exception of a narrow region
close to the $\mh \simeq 2\, m_S$ threshold \cite{BDV}.

There is only one case in which the effective approximation does not
work sufficiently well, namely when the bottom contribution becomes
very relevant. This can happen in the MSSM when $\tan\beta$, i.e.~the
ratio of the vacuum expectation values (vev) of the neutral components
of the two Higgs doublets, becomes large. In such a situation, in
principle, the exact computation of the NLO bottom contribution to the
cross section should be employed. However, explicit analytic results
for the part of the NLO bottom contribution that is induced by
two-loop bottom-sbottom-gluino diagrams have not been made available
so far.

In this paper we present an approximate evaluation of the
bottom-sbottom-gluino diagrams, based on an asymptotic expansion in
the large supersymmetric masses that is valid up to and including
terms of ${\cal O}(\mb^2/m_\phi^2)$, ${\cal O}(\mb/M)$ and ${\cal
  O}(\mz^2/M^2)$, where $m_\phi$ denotes a Higgs-boson mass and $M$
denotes a generic superparticle mass ($M=\mg,m_{\sbu},m_{\sbd}$).
Together with the known exact results for the (s)bottom-gluon diagrams
\cite{SDGZ,babis1,ABDV,MS,BDV}, our new result allows us to obtain
effective $K$ factors that can be easily implemented in computer codes
to provide an accurate and efficient evaluation of the cross section
for Higgs boson production in the MSSM. Such $K$ factors are expected
to be at the same level of accuracy as in the SM, i.e.~within a few
per cent of the exact value for the light Higgs and up to ten per cent
for the heavy Higgs, with however a much better accuracy expected if
the heavy-Higgs mass is below all the heavy-particle thresholds.

The paper is organized as follows: in section \ref{sec:general} we
summarize general results on the cross section for Higgs boson
production via gluon fusion.  In section \ref{sec:exp} we outline the
way we perform the asymptotic expansion of the bottom-sbottom-gluino
diagrams.  Section \ref{sec:2loopres} contains the explicit results
for the NLO bottom contribution both in $\drbar$ and in the on-shell
(OS) scheme.  In section \ref{sec:num} we discuss the numerical
relevance of the NLO bottom corrections devoting particular attention
to the interplay between the computation of the Higgs mass and that of
its production cross section. In the last section we discuss an
approximate way to take the NLO bottom contribution into account,
based on an improved LO term.  Finally, in appendix A we specialize to
the MSSM case the general exact results for the real radiation derived
in ref.~\cite{BDV}, while appendix B contains the explicit expressions
for the shifts from the $\drbar$ to the OS parameters in the sbottom
sector.


\vspace*{1mm}
\section{Higgs boson production via gluon fusion at NLO 
  in the MSSM }
\label{sec:general}

In this section we recall for completeness some general results on
Higgs boson production via gluon fusion.  The hadronic cross section
for Higgs boson production at center-of-mass energy $\sqrt{s}$ can be
written as
\be
\sigma(h_1 + h_2 \to \phi+X)  \,=\, 
          \sum_{a,b}\int_0^1 dx_1 dx_2 \,\,f_{a,h_1}(x_1,\muF)\,
         f_{b,h_2}(x_2,\muF)  \times
\int_0^1 dz~ \delta \left(z-\frac{\tau_\phi}{x_1 x_2} \right)
\hat\sigma_{ab}(z)~,
\label{sigmafull}
\ee
where $\phi=(h,H)$, $\tau_\phi= m^2_\phi/s$, $\muF$ is the
factorization scale, $f_{a,h_i}(x,\muF)$ the parton density of the
colliding hadron $h_i$ for the parton of type $a$ (for $a = g,q,\bar{q}$),
and $\hat\sigma_{ab}$ the cross section for the partonic subprocess $
ab \to \phi +X$ at the center-of-mass energy $\hat{s}=x_1 \,x_2\,
s=m^2_\phi/z$. The latter can be written in terms of the LO
contribution $\sigma^{(0)}$ and a coefficient function $G_{ab}(z)$
as
\be
\hat\sigma_{ab}(z)=
\sigma^{(0)}\,z \, G_{ab}(z) \, .
\label{Geq}
\ee
 
We consider now the production of the lightest CP-even Higgs boson,
$h$, through gluon fusion\footnote{ For the heaviest eigenstate, $H$,
  general formulae for the production cross section can be obtained
  straightforwardly with the replacements ($ \sin\alpha \rightarrow
  -\cos\alpha,\,\cos\alpha \rightarrow \sin\alpha$) in
  eqs.~(\ref{ggh}) and (\ref{real}) and in appendix A.}. The LO term
can be written as
\be
\sigma^{(0)}  =  
\frac{G_\mu \,\alpha_s^2 (\muR)  }{128\, \sqrt{2} \, \pi}\,
\left|T_F \left( -\sin\alpha \,{\mathcal H}^{1\ell}_1 +
\cos\alpha \,{\mathcal H}^{1\ell}_2 \right)  \right|^2~,
\label{ggh}
\ee
where $G_\mu$ is the muon decay constant, $\alpha_s(\muR)$ is the
strong gauge coupling expressed in the $\overline{\rm MS}$
renormalization scheme at the scale $\muR$, $T_F=1/2$ is a color
factor, and $\alpha$ is the mixing angle in the CP-even Higgs sector
of the MSSM. ${\mathcal H}_i$ ($i = 1,2$) are the form factors for the
coupling of the neutral, CP-even component of the Higgs doublet $H_i$
with two gluons, which we decompose in one- and two-loop parts as
\be
{\mathcal H}_i ~=~ {\mathcal H}_i^{1\ell}
~+~ \frac{\alpha_s}{\pi} \,  {\mathcal H}_i^{2\ell}
~+~{\cal O}(\alpha_s^2)~.
\label{Hdec}
\ee
The one-loop form factors ${\cal H}_1^{1\ell}$ and ${\cal
  H}_2^{1\ell}$ contain contributions from diagrams involving quarks
or squarks. The two-loop form factors ${\cal H}_1^{2\ell}$ and ${\cal
  H}_2^{2\ell}$ contain contributions from diagrams involving quarks,
squarks, gluons and gluinos. Focusing on the contributions involving
the third-generation quarks and squarks, and exploiting the structure
of the Higgs-quark-quark and Higgs-squark-squark couplings, the form
factors ${\mathcal H}_i$ can be written to all orders in the strong
interactions as \cite{DS}
\bea
{\mathcal H}_1 & = & \lambda_t \,\left[
\mt \,\mu\,\sdt\,F_t
\,+ \mz^2 \,\sdbeta \,D_t  \right] \;+ 
\lambda_b \,\left[\mb\,A_b\,\sdb\,F_b \,+  2\,\mb^2\,G_b \,+ 
2\, \mz^2 \,c_\beta^2 \,D_b
\right]\,, \label{eq:H1} \\
{\mathcal H}_2 & = & \lambda_b\,\left[
\mb \,\mu\,\sdb\,F_b
\,-\mz^2 \,\sdbeta \,D_b  \right] + 
\lambda_t\, \left[
\mt\,A_t\,\sdt\,F_t \,+  2\,\mt^2\,G_t \,- 
2\, \mz^2 \,s_\beta^2 \,D_t
\right]\label{eq:H2}~.
\eea
In  the equations  above $\lambda_t  = 1/\sin\beta$  and  $\lambda_b =
1/\cos\beta$, where $\tan\beta \equiv v_2/v_1$ is the ratio of the vev
of the two Higgs doublets.  Also, $\mu$ is the higgsino mass parameter
in  the  MSSM  superpotential,   $A_q$  (for  $q=t,b$)  are  the  soft
SUSY-breaking  Higgs-squark-squark couplings  and  $\theta_q$ are  the
left-right  squark  mixing angles  (here  and  thereafter  we use  the
notation   $s_\varphi   \equiv   \sin\varphi,  \,   c_\varphi   \equiv
\cos\varphi$ for  a generic angle $\varphi$). The  functions $F_q$ and
$G_q$  appearing in  eqs.~(\ref{eq:H1}) and  (\ref{eq:H2})  denote the
contributions  controlled by  the  third-generation Yukawa  couplings,
while $D_q$  denotes the  contribution controlled by  the electroweak,
D-term-induced  Higgs-squark-squark  couplings.   The  latter  can  be
decomposed as
\be
D_q  = \frac{I_{3q}}2 \, \widetilde{G}_q
+ c_{2\theta_{\tilde q}} \, 
\left(\frac{I_{3q}}2 -  Q_q \,s^2_{\theta_\smallw} 
\right) \,\widetilde{F}_q \, , \label{eq:Dq}
\ee
where $I_{3q}$ denotes the third component of the electroweak isospin
of the quark $q$, $Q_q$ is the electric charge and $\theta_\smallw$ is
the Weinberg angle.

The one-loop functions entering ${\cal H}_1^{1\ell}$ and ${\cal
  H}_2^{1\ell}$ are:
\bea
F_q^{1\ell} ~=~ \widetilde F_q^{1\ell}
& =&  \frac{1}{2}\,\left[
\frac1{m^2_{\tilde{q}_{1}}} {\mathcal G}^{1\ell}_{0}(\tau_{\tilde{q}_{1}}) -
\frac1{m^2_{\tilde{q}_{2}}} {\mathcal G}^{1\ell}_{0}(\tau_{\tilde{q}_{2}}) 
\right]\, , \label{eq:F1l}\\
\nn\\
G_q^{1\ell} & =& \frac{1}{2}\,\left[
\frac1{m^2_{\tilde{q}_{1}}} {\mathcal G}^{1\ell}_{0}(\tau_{\tilde{q}_{1}}) +
\frac1{m^2_{\tilde{q}_{2}}} {\mathcal G}^{1\ell}_{0} (\tau_{\tilde{q}_{2}}) + 
\frac1{m_q^2} {\mathcal G}^{1\ell}_{1/2} (\tau_q)\right]~, \label{eq:G1l}\\
\widetilde G_q^{1\ell} & =& \frac{1}{2}\,\left[
\frac1{m^2_{\tilde{q}_{1}}} {\mathcal G}^{1\ell}_{0} (\tau_{\tilde{q}_{1}}) +
\frac1{m^2_{\tilde{q}_{2}}} {\mathcal G}^{1\ell}_{0} (\tau_{\tilde{q}_{2}}) 
\label{eq:Gtilde}\right]~,
\eea
where $\tau_k \equiv 4\,m_k^2/m_h^2$, and the functions ${\mathcal
  G}^{1\ell}_{0}$ and ${\mathcal G}^{1\ell}_{1/2}$ read
\bea
{\mathcal G}^{1\ell}_{0} (\tau) & =& ~~~~\,\tau \!\left[ 1 + \frac{\tau}{4}\, 
 \ln^2 \left(\frac{\sqrt{1- \tau} - 1}{\sqrt{1- \tau} + 1}\right) \right]\,,
\label{eq:4} \\
{\mathcal G}^{1\ell}_{1/2} (\tau) & = & - 2\,\tau
 \left[ 1 - \frac{ 1 -\tau}4  \,   
 \ln^2 
  \left(\frac{\sqrt{1-\tau} - 1}{\sqrt{1-\tau} + 1} \right) \right] \,.
\label{eq:3}
\eea
The analytic continuations are obtained with the replacement $m_h^2
\rightarrow m_h^2 + i \epsilon$~. For later convenience, we recall the
behavior of ${\mathcal G}^{1\ell}_{0}$ and ${\mathcal
  G}^{1\ell}_{1/2}$ in the limit in which the Higgs boson mass is much
smaller or much larger than the mass of the particle running in the
loop.  In the first case, i.e.~$\tau\gg 1$, which applies to the top
and squark contributions for the light-Higgs case,
\be
{\mathcal G}^{1\ell}_{0} \rightarrow -\frac13 -\frac{8}{45\,\tau} 
~+~{\cal O}(\tau^{-2})~,~~~~~~~~~~
{\mathcal G}^{1\ell}_{1/2} \rightarrow -\frac43-\frac{14}{45\,\tau} 
~+~{\cal O}(\tau^{-2})~,
\label{Glimit}
\ee
while in the opposite case, i.e.~$\tau\ll 1$, which is relevant for
the bottom quark,
\be
{\mathcal G}^{1\ell}_{0} \rightarrow \tau
~+~{\cal O}(\tau^{2})~,~~~~~~~~~~
{\mathcal G}^{1\ell}_{1/2} \rightarrow - 2\,\tau  +\frac{\tau}2 
 \ln^2 (\frac{-4}{\tau}) ~+~{\cal O}(\tau^{2})~.
\label{GLimit}
\ee

The coefficient function $G_{ab}(z)$ in eq.~(\ref{Geq}) can be
decomposed, up to NLO terms, as
\be
G_{a b}(z)  ~=~  G_{a b}^{(0)}(z) 
~+~ \frac{\alpha_s}{\pi} \, G_{a b}^{(1)}(z) ~+~{\cal O}(\alpha_s^2)\, ,
\label{Gdec}
\ee
with the LO contribution given only by the gluon-fusion channel:
\bea
G_{a b}^{(0)}(z) & = & \delta(1-z) \,\delta_{ag}\, \delta_{bg} \, .
\eea
The NLO terms include, besides the $gg$ channel, also the one-loop
induced processes $gq \rightarrow qh$ and $q \bar{q} \rightarrow g h$:
\bea
\label{ggg}
G_{g g}^{(1)}(z) & = & \delta(1-z) \left[C_A \, \frac{~\pi^2}3 
\,+ \,\beta_0 \, \ln \left( \frac{\muR^2}{\muF^2} \right) 
 \,+ \,2\,{\rm Re}\left(\frac{-\sin\alpha \,{\mathcal H}^{2\ell}_1 
+\cos\alpha \,{\mathcal H}^{2\ell}_2}{
-\sin\alpha \,{\mathcal H}^{1\ell}_1 
+ \cos\alpha \,{\mathcal H}^{1\ell}_2 } \right)  \right]  \nn \\
&+ &  P_{gg} (z)\,\ln \left( \frac{\hat{s}}{\muF^2}\right) +
    C_A\, \frac4z \,(1-z+z^2)^2 \,{\cal D}_1(z) +  C_A\, {\cal R}_{gg}  \, , 
\label{real}
\eea

\be
G_{q \bar{q}}^{(1)}(z) ~=~   {\cal R}_{q \bar{q}} \, , ~~~~~~~~~~~
G_{q g}^{(1)}(z) ~=~  P_{gq}(z) \left[ \ln(1-z) + 
 \frac12 \ln \left( \frac{\hat{s}}{\muF^2}\right) \right] + {\cal R}_{qg} \,,
\label{qqqg}
\ee
where the LO Altarelli-Parisi splitting functions are
\be
P_{gg} (z) ~=~2\,  C_A\,\left[ {\cal D}_0(z) +\frac1z -2 + z(1-z) \right]
\label{Pgg} \, ,~~~~~~~~~~~
P_{gq} (z) ~=~  C_F \,\frac{1 + (1-z)^2}z~. 
\ee
In the equations above, $C_A =N_c$ and $C_F = (N_c^2-1)/(2\,N_c)$
($N_c$ being the number of colors), $\beta_0 = (11\, C_A - 2\, N_f)/6
$ ($N_f$ being the number of active flavors) is the one-loop
$\beta$-function of the strong coupling in the SM, and
\be
{\cal D}_i (z) =  \left[ \frac{\ln^i (1-z)}{1-z} \right]_+  \label {Dfun} \, .
\ee

The $gg$-channel contribution, eq.~(\ref{real}), involves two-loop
virtual corrections to $g g \rightarrow h$ and one-loop real
corrections from $ gg \to h g$. The former, regularized by the
infrared-singular part of the real emission cross section, are
displayed in the first line of eq.~(\ref{real}). The second line
contains the non-singular contribution from the real gluon emission in
the gluon fusion process.  The latter contribution as well as the ones
due to the $ q \bar q \to h g $ annihilation channel and the
quark-gluon scattering channel, eq.~(\ref{qqqg}), are obtained from
one-loop diagrams where only quarks or squarks circulate in the
loop. General expressions for the functions ${\cal R}_{gg},\, {\cal
  R}_{q \bar q},\, {\cal R}_{q g}$ can be found in ref.~\cite{BDV}
(see also refs.~\cite{ehsv,BG}). In appendix A we provide expressions
in which the contribution of the bottom quark is kept exact while
those of the top quark and of the squarks are evaluated in the limit
of vanishing Higgs mass.

The two-loop top/stop contributions to the form factors ${\cal
  H}_{1,2}^{2\ell}$ entering eq.~(\ref{ggg}) are fully under control
in the light-Higgs case. Typically, the mass ratios between the Higgs
and the particles running in the loops allow for the evaluation of the
relevant diagrams via a Taylor expansion in the Higgs mass, with the
zero-order term in the series already a very good approximation of the
full result.  The case of the two-loop bottom/sbottom contributions is
obviously different.  In general, Taylor-expanded evaluations of the
relevant diagrams are no longer viable, due to the presence of the
light bottom quark in the loops. Thus, the diagrams must be evaluated
either exactly or via an asymptotic expansion in a large mass or
momentum.

In the following sections we present the result for the NLO bottom
contribution, combining earlier results in the literature with our new
calculation of the bottom-sbottom-gluino contribution. The latter has
been obtained via an asymptotic expansion, retaining terms of ${\cal
  O}(\mb^2/m_h^2)$, ${\cal O}(\mb/M)$ and ${\cal O}(\mz^2/M^2)$.


\section{Outline of the calculation}
\label{sec:exp}

An exact analytic evaluation of the bottom-sbottom-gluino contribution
to ${\mathcal H}^{2\ell}_{1,2}$ is, at the moment, beyond our
computational ability. However, it is reasonable to assume that all of
the supersymmetric particles are much heavier than the lightest Higgs
boson and the bottom quark, and look for an approximate evaluation of
the diagrams in terms of a small-momentum (large-mass) expansion. We
follow this path by performing a large-mass expansion, assuming all
the supersymmetric particles to be heavy but without requiring any
specific hierarchy among them.

After generating the two-loop diagrams involving bottom, sbottom and
gluino that contribute to the process $g(q_1) + g(q_2) \to h(q)$ with
the help of {\tt FeynArts} \cite{feynarts}, we separate them in two
classes: {\it i)} those that can be evaluated via an ordinary Taylor
expansion in powers of $q^2/M^2$, of which we keep only the term of
order zero; {\it ii)} the diagrams that require an asymptotic
expansion.  We recall that a Taylor expansion of a two-loop diagram in
the external momentum $q^2$ is viable for values of $q^2$ up to the
first physical threshold. In our case, diagrams with a physical
threshold at $q^2 = 4\, \mb^2$, when Taylor-expanded in $q^2$, exhibit
an infrared (IR) divergent behavior as the bottom mass is sent to
zero.  Thus, these diagrams belong to the class {\it ii}.
\begin{figure}[t]
\begin{center}
\epsfig{figure=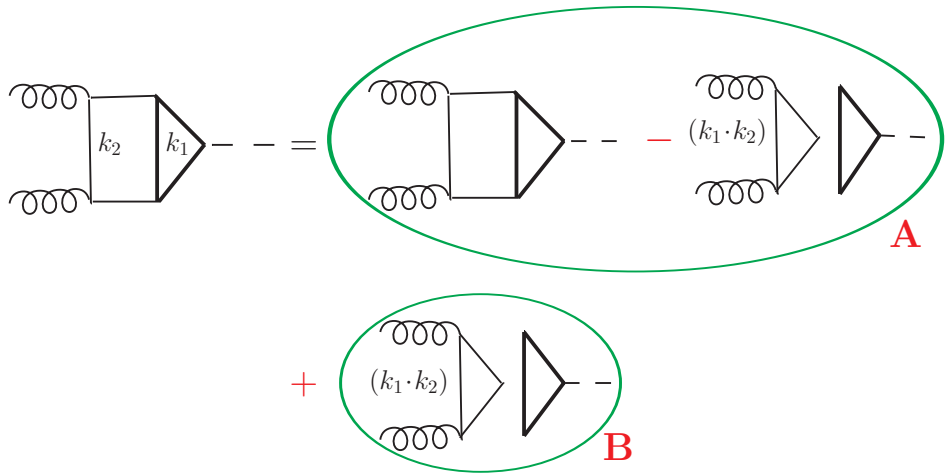,width=13cm}
\end{center}
\caption{Pictorial example of the asymptotic expansion of a two-loop
  diagram containing one subintegration ($k_1$) with only heavy
  particles (bold lines) and the other ($k_2$) with light
  particles. To the original diagram its IR-divergent part,
  represented by the disconnected diagram, is subtracted and added
  forming the contribution that can be evaluated via a Taylor
  expansion (A) and the one that should evaluated exactly (B). See
  text for a detailed explanation. }
\label{fig1}
\end{figure}

Class-{\it i} diagrams are expressed in terms of two-loop vacuum
integrals that can be evaluated using the results of
ref.~\cite{tausk}. Concerning the diagrams belonging to class {\it
  ii}, reviews of the method of asymptotic expansions of Feynman
diagrams with respect to masses or momenta can be found in
ref.~\cite{Asy}. In practice, we generate the expansion of a diagram
by adding and subtracting to it the part of the diagram itself that
becomes IR-divergent when $\mb$ and $q^2$ are sent to zero.  Formally
we are adding nothing to the original diagram but, as
graphically\footnote{The diagrams have been drawn using JaxoDraw
  \cite{bt}.} exemplified in fig.~\ref{fig1}, this construction allows
us to separate the diagram in two parts: part A in fig.~\ref{fig1}
which, being by construction IR-safe, can be evaluated via a Taylor
expansion in the same way as class-{\it i} diagrams; part B in
fig.~\ref{fig1}, containing the IR-divergent contribution, which
should be evaluated exactly.

The IR-divergent part of a diagram is constructed in the following
way. We first note that in all the diagrams entering our calculation
one can choose a routing of momenta such that the connecting
propagators, i.e.~the propagators that contain both integration
momenta $k_1$ and $k_2$, are always accompanied by a heavy mass
$M$. Furthermore, only one subintegration, let us assume the one on
$k_2$, is IR divergent. Then, one can rewrite the connecting
propagators using the identity
\be
\frac1{(k_1+k_2)^2-M^2} = \frac1{k_1^2-M^2} -
 \frac{k_2^2 +2 \,k_1\cdot k_2}{[(k_1+k_2)^2-M^2](k_1^2-M^2)}~.
\label{idint}
\ee
The first term on the r.h.s.~of eq.~(\ref{idint}) leads to a
disconnected integral (product of two one-loop integrals) that
contains the IR-divergent contributions present in the original
diagram.  This term can be evaluated exactly, i.e.~for arbitrary
$q^2$, giving rise to the $\ln( q^2/\mb^2)$ terms that describe the
physical threshold.  The second term, instead, leads to a two-loop
integral with improved infrared convergence in the $k_2$ integration
and improved ultraviolet convergence in the $k_1$ integration.
Therefore, if, for example, the original integral is logarithmically
IR divergent in the $k_2$ integration when $q \to 0$ and $\mb \to 0$,
the corresponding two-loop integral associated with the second term in
eq.~(\ref{idint}) evaluated at $q^2=\mb =0$ is no longer IR divergent,
but it actually gives a finite result that differs from the result
valid for $q^2 \neq 0$ and $\mb \neq 0$ by terms of ${\cal O}(m^2 /M^2
\ln (m^2/M^2))$, where $m^2$ denotes either $q^2$ or $m_b^2$. In
general, a repeated application of eq.~(\ref{idint}), controlled by
the power counting in the IR-divergent terms, allows us to construct
the IR-divergent part of any diagram in terms of products of one-loop
integrals with numerators that contain terms of the form $(k_{i}\cdot
q_{j})^m, \, (k_i \cdot k_j)^n\,(i,j =1,2)$ where $m,\, n$ are generic
powers. The Passarino-Veltman reduction method is then applied to
eliminate the numerators and express the result in terms of the known
one-loop scalar integrals \cite{PV}. A check of the validity of our
construction of the IR-divergent part of a diagram is given by the
evaluation of its part A. Indeed, one verifies explicitly that the
IR-divergent contributions of the original diagram are canceled by the
terms constructed via eq.~(\ref{idint}), so that the final result for
part A is free of any $\ln( q^2/\mb^2)$ or $q^2/\mb^2$ term.


\section{Two-loop bottom/sbottom contributions}
\label{sec:2loopres}

In this section we present the result for the two-loop bottom/sbottom
contribution to the form factor for Higgs boson production via gluon
fusion. We stress that, in the MSSM, the result for the production
cross section of a Higgs boson is strictly linked to the computation
of its mass, i.e., both observables should be computed in terms of the
same set of SUSY parameters, defined in the same way beyond tree
level. There is however an important difference between the two
calculations. In the computation of the one-loop corrections to the
Higgs masses, the diagrams involving the bottom quark are suppressed
by the bottom mass and can be safely neglected, resulting in a
one-loop contribution that is actually due to the sbottom diagrams
only. This implies that in the two-loop calculation of the Higgs
masses the only couplings that require a one-loop renormalization are
the trilinear sbottom-Higgs couplings, while the definition of the
bottom-Higgs Yukawa coupling beyond tree level is irrelevant.  On the
other hand, in the one-loop calculation of the amplitude for Higgs
production both the bottom-Higgs Yukawa coupling and the trilinear
sbottom-Higgs couplings play a role, thus they both require a one-loop
definition when the two-loop contributions are computed.

\begin{figure}[ht]
\begin{center}
\mbox{
\epsfig{figure=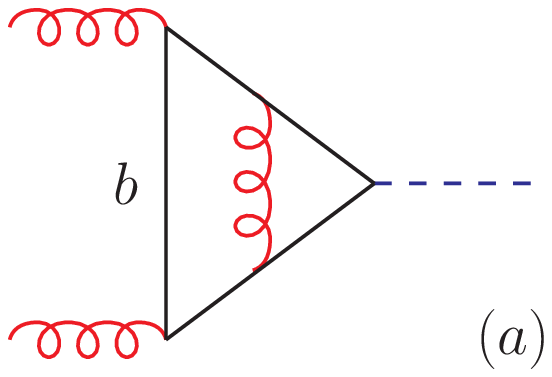,width=4.3cm}~~~~~~
\epsfig{figure=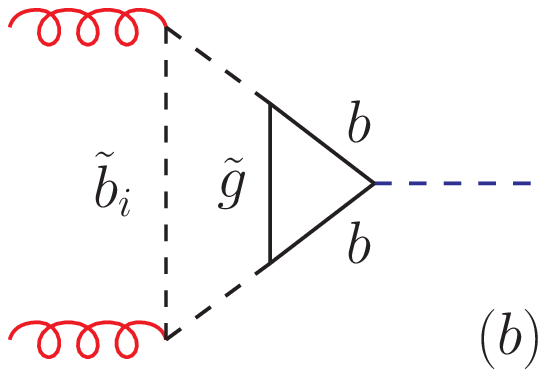,width=4.3cm}
}
\caption{Examples of two-loop diagrams involving the Higgs-bottom coupling.}
\label{fig:diaghq}
\end{center}
\end{figure}

\begin{figure}[ht]
\begin{center}
\mbox{
\epsfig{figure=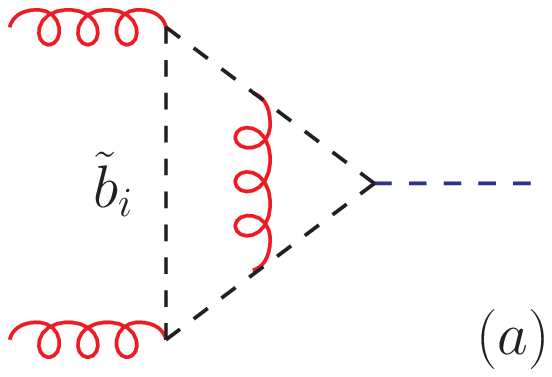,width=4.3cm}~~~~~~
\epsfig{figure=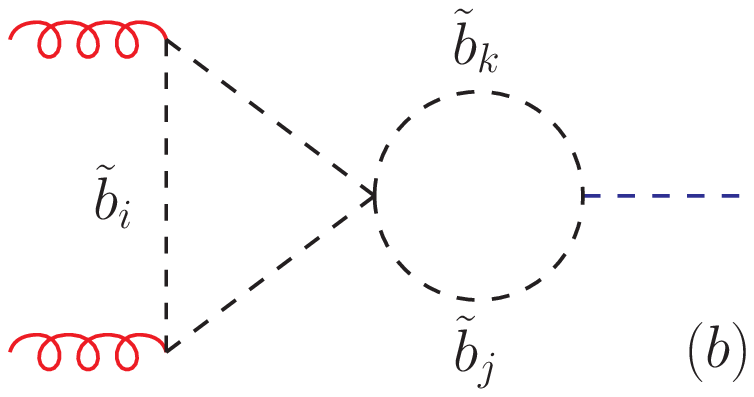,width=5.7cm}~~~~~~
\epsfig{figure=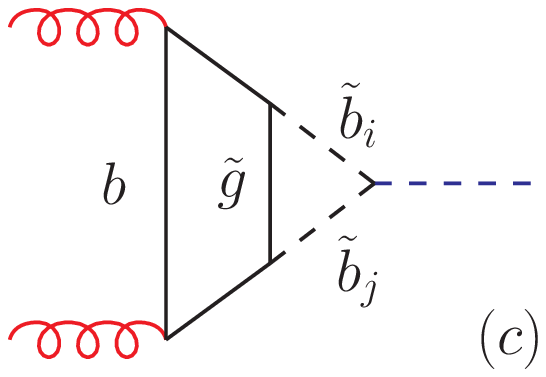,width=4.3cm}
}
\caption{Examples of two-loop diagrams involving the Higgs-sbottom
  coupling.}
\label{fig:diaghsq}
\end{center}
\end{figure}
  
In the following we will discuss separately the contributions to the
two-loop form factors ${\mathcal H}^{2\ell}_{1,2}$ from the diagrams
involving the Higgs-bottom coupling (examples of which are shown in
fig.~\ref{fig:diaghq}) and those from the diagrams involving the
Higgs-sbottom couplings (see fig.~\ref{fig:diaghsq}).  In our
approximation of neglecting terms beyond ${\cal O}(m_b/M)$, the former
contribute only to the function $G_b$ in eq.~(\ref{eq:H1}), while the
latter contribute only to the functions $F_b,\, \widetilde F_b$ and
$\widetilde G_b$ in eqs.~(\ref{eq:H1}) and (\ref{eq:H2}). For both
kinds of contributions, we first report the results obtained in the
$\drbar$ renormalization scheme, which is the scheme employed by
several public computer codes \cite{mssmcodes} that return the MSSM
mass spectrum starting from a set of high-energy boundary conditions
for the SUSY-breaking parameters.  We then discuss how to move from
the $\drbar$ scheme to a different renormalization prescription, which
generalizes the one introduced for the calculation of the Higgs masses
in ref.~\cite{bdsz2} and implemented in the code {\tt FeynHiggs}
\cite{feynhiggs}.  With a slight abuse of language, since some of the
amplitudes involved in the renormalization of the SUSY parameters are
in fact evaluated off mass shell, we refer to this prescription as to
the ``on-shell'' scheme.

\subsection{Contributions controlled by the Higgs-bottom 
coupling}
\label{sec:botcontr}

We start by considering the contributions of the two-loop diagrams
involving the Higgs-bottom coupling.  In our approximation those are
the only diagrams that contribute to the function $G_b$ in
eq.~(\ref{eq:H1}). The two-loop part of the function can be
decomposed as
\be
G_b^{2\ell} ~=~ C_F\,\left(G_b^{(g,C_{F})} + G_b^{(\gl,C_{F})}\right) 
~+~C_A\,\left(G_b^{(g,C_{A})} + G_b^{(\gl,C_{A})}\right)~.
\ee
Assuming that the one-loop form factor ${\mathcal H}_{1}^{1\ell}$ is
expressed in terms of $\drbar$-renormalized parameters evaluated at
the scale $Q^2$, the contribution of the two-loop diagrams with bottom
quarks and gluons (fig.~\ref{fig:diaghq}a) reads:
\be
2\,\mb^2\,G_b^{(g,C_{F})} ~=~ {\cal F}^{(2\ell,a)}_{1/2}(\tau_b) ~+~
{\cal F}^{(2\ell,b)}_{1/2}(\tau_b)\,\left(\ln\frac{\mb^2}{Q^2}-
\frac13\right)~,~~~~~~
2\,\mb^2\,G_b^{(g,C_{A})} ~=~ {\cal G}^{(2\ell,C_A)}_{1/2}(\tau_b)~.
\label{gdl}
\ee
Exact expressions for the functions ${\cal F}^{(2\ell,a)}_{1/2}$,
${\cal F}^{(2\ell,b)}_{1/2}$ and ${\cal G}^{(2\ell,C_A)}_{1/2}$ are
given in eqs.~(2.12), (2.13) and (3.8) of ref.~\cite{ABDV},
respectively. In the limit $\tau \ll 1$ they reduce to
\bea
{\cal F}^{(2\ell,a)}_{1/2}(\tau) & = & -\tau\,\biggr[
9 + \frac{9}{5}\,\zeta_2^2 - \zeta_3-(1+\zeta_2+ 4\,\zeta_3)\,
\ln(\frac{-4}{\tau})- (1-\zeta_2)\,\ln^2(\frac{-4}{\tau})\nonumber\\
&&~~~~~~+\frac{1}{4}\,\ln^3(\frac{-4}{\tau})
+\frac{1}{48}\,\ln^4(\frac{-4}{\tau})\biggr] ~+~ {\cal O}(\tau^2)~,\\
&&\nonumber\\
{\cal F}^{(2\ell,b)}_{1/2}(\tau)  &=& 3\,\tau\,\biggr[
1 + \frac12\,\ln(\frac{-4}{\tau})
- \frac14\ln^2(\frac{-4}{\tau})\biggr]  ~+~ {\cal O}(\tau^2)~,\\
&&\nonumber\\
{\cal G}^{(2\ell,C_A)}_{1/2}(\tau) & = & -\tau\,\biggr[
3 - \frac{8}5\,\zeta_2^2 - 3\,\zeta_3 + 3\,\zeta_3\,\ln(\frac{-4}{\tau})
-\frac14\,(1+2\,\zeta_2)\,\ln^2(\frac{-4}{\tau})\nonumber\\
&&~~~~~~
-\frac1{48}\,\ln^4(\frac{-4}{\tau})\biggr] ~+~ {\cal O}(\tau^2)~,
\eea
where $\zeta_2$ and $\zeta_3$ are Riemann's zeta functions. 

The contributions of the two-loop diagrams with bottom, sbottom and
gluino (fig.~\ref{fig:diaghq}b) require a dedicated calculation. Up to
and including terms of ${\cal O}(m_b^2/m_h^2)$ and ${\cal O}(\mb/M)$,
and assuming that ${\mathcal H}_{1}^{1\ell}$ is expressed in terms of
$\drbar$-renormalized parameters evaluated at the scale $Q^2$, they
read:
\bea 
2\, m_b^2 \,G_b^{(\gl,C_{F})} &=&
\frac43\, {\cal F}^{(2\ell,b)}_{1/2}(\tau_b)\,{\frac{
(\delta m_b)}{m_b}}^{\scriptscriptstyle SUSY}\!\!\!
-~ \frac14\, {\mathcal G}^{1\ell}_{1/2} (\tau_b)\,\frac{\mg}{m_b}\,\sdb  \,
\left(\frac{\x1g}{1-\x1g} \ln \x1g-\frac{x_2}{1-x_2}\ln x_2\right)\nn\\
&-&  \frac{m_b}{\mg}\,\sdb\, \biggr\{ \frac1{6\, \x1g\, (1-\x1g)^3}\biggr[ 
 (1-\x1g)^3\, \ln\frac{\mg^2}{Q^2} 
+2 \left( \x1g^3 + 2\, \x1g^2  \right)\, \ln \x1g  \nn\\ 
&& 
- 3\, \left( \x1g^3 - \x1g -2\, \x1g^2 \,\ln \x1g \right) 
\,\ln (\frac{-m_h^2}{\mg^2} )  +
5\, \x1g^3 - 5\, \x1g^2 + \x1g -1 \nn\\
&& \left. -12\, \x1g^2 \,{\rm Li_2}\left(1-\frac1{\x1g} \right) -
   6\, \x1g^2\, \ln^2 \x1g \right] ~~-~~(x_1\rightarrow x_2)~\biggr\}~,
\label{YCFG} \\
\nn \\
2 \,m_b^2 \, G_b^{(\gl,C_{A})} &=& 
\frac{m_b}{\mg}\,\sdb \, \biggr\{ \frac1{6 \, (1-\x1g)^2}\left[
2 \, \x1g\, (1+\x1g) \,\ln \x1g + 2\,\x1g -2  - 
6\, \x1g\,  {\rm Li}_2\left(1-\frac1{\x1g} \right)  \right.
\nn\\
&& \left. - 3\, \x1g\, \ln^2 \x1g + 3\,(1-\x1g + \x1g\, \ln \x1g) \,
\ln (\frac{-m_h^2}{\mg^2}) \right] ~~-~~(x_1\rightarrow x_2)~\biggr\}~,
\label{YCAG} 
\eea
where $x_i = m_{\tilde b_i}^2 /\mg^2\,$, and $(\delta
m_b)^{\scriptscriptstyle SUSY}$ denotes the SUSY contribution to the
bottom self-energy, in units of $C_F\,\alpha_s/\pi$ and in the limit
of vanishing $m_b$:
\be
\label{dmbsusy}
\frac{(\delta m_b)}{\mb}^{\scriptscriptstyle SUSY} ~=~ -\frac14\,\left[
\ln\frac{\mg^2}{Q^2}  + f(x_1)+f(x_2) + \frac{\mg}{\mb}\,\sdb\,
\left(\frac{\x1g}{1-\x1g} \ln \x1g-\frac{x_2}{1-x_2}\ln x_2\right)\right]~,
\ee
where 
\be
f(x) ~=~ \frac{ x - 3}{4\, (1-x)} + \frac{x\,(x-2)}{2\,(1-x)^2} \,\ln x ~.
\ee

If the bottom-quark contribution to ${\mathcal H}_{1}^{1\ell}$ is
expressed in terms of the pole bottom mass $M_b$, the functions
$G_b^{(g,C_{F})}$ and $G_b^{(\tilde g,C_{F})}$ are shifted with respect
to their expressions in eqs.~(\ref{gdl}) and (\ref{YCFG}). In
particular, the former becomes
\be
\label{GbOS}
2\,\mb^2\,G_b^{(g,C_{F})} ~=~ {\cal F}^{(2\ell,a)}_{1/2}(\tau_b) ~+~
\frac43 \, {\cal F}^{(2\ell,b)}_{1/2}(\tau_b)~,
\ee
and the term proportional to $(\delta m_b)^{\scriptscriptstyle SUSY}$
in the first line of eq.~(\ref{YCFG}) is canceled out.

Eqs.~(\ref{YCFG}) and (\ref{dmbsusy}) show that the
bottom-sbottom-gluino contribution to ${\mathcal H}_{1}^{2\ell}$
contains terms enhanced by the large ratio $\mg/\mb$. Recalling the
definition of $\tau_b$, it is clear that those terms are in fact of
${\cal O}(\mb\,\mg/m_h^2)\,$, i.e., they still vanish as
$\mb\rightarrow 0$ but they are enhanced by the ratio $\mg/m_h$. Such
terms arise from two-loop diagrams in which the helicity flip on the
fermion loop is achieved via a gluino mass insertion instead of a
bottom mass insertion, and they by far dominate the new-physics
contribution to the two-loop part of the form factors, no matter
whether the bottom-quark contribution to ${\mathcal H}_{1}^{1\ell}$ is
expressed in terms of the pole bottom mass $M_b$ or in terms of the
$\drbar$-renormalized bottom mass $\widehat \mb$. However, we notice
that all of the two-loop ${\cal O}(\mb\,\mg/m_h^2)$ terms cancel out
if the one-loop bottom contribution to the function $G_b^{1\ell}$ is
computed in terms of $M_b$, but the function itself is multiplied by
$\widehat \mb\,M_b$ instead of $M_b^2$. As a result the function
$G_b^{2\ell}$ is further shifted, with respect to the expression
corresponding to the use of $M_b$ in the whole one-loop contribution,
by
\be
\label{shift}
2\,\mb^2\,G_b^{2\ell} ~\longrightarrow~ 2\,\mb^2\,G_b^{2\ell} ~-~
{\mathcal G}^{1\ell}_{1/2} (\tau_b)\, C_F
\left[ \frac34\,\ln \frac{\mb^2}{Q^2} -\frac54 +
\frac{(\delta m_b)}{\mb}^{\scriptscriptstyle SUSY}\right]~.
\ee

This manipulation amounts to differentiating, in the one-loop
contribution, between the parameter that describes the mass of the
bottom quark running in the loop -- which is identified with $M_b$ --
and the parameter that describes the Yukawa coupling of the bottom
quark to the Higgs boson -- which is identified with $\widehat
\mb$. We recall that, in the MSSM, the running bottom mass $\widehat
\mb$ can be related to the corresponding SM parameter $\overline \mb$
as \cite{hrs}
\be
\label{mbrun}
\widehat \mb ~=~ \frac{\overline \mb\,(1+\delta_b)}{1+\epsilon_b\,\tan\beta}~,
\ee
where $\delta_b$ denotes terms that are not enhanced by $\tan\beta$
and, to ${\cal O}(\alpha_s)$,
\be
\label{epsilon}
\epsilon_b ~=~ \frac{\alpha_s\,C_F}{4\pi}\,\frac{2\,\mu\,\mg}{\bu-\bd}\,
\left(\frac{\x1g}{1-\x1g} \ln \x1g
-\frac{x_2}{1-x_2}\ln x_2\right)~.
\ee
Since $\sdb = 2\,\mb\,(A_b+\mu\tan\beta)/(\bu-\bd)\,$, it is easy to
see that the terms enhanced by $\mg/\mb$ in $G_b^{(\tilde g,C_{F})}$
do indeed contain $\epsilon_b\,\tan\beta$. In the effective-theory
language of ref.~\cite{resum} we can argue that, by expressing the
bottom Yukawa coupling entering ${\mathcal H}_{1}^{1\ell}$ in terms of
$\widehat \mb$ as defined in eq.~(\ref{mbrun}), we ``resum'' in the
one-loop part of the form factor the $\tan\beta$-enhanced threshold
corrections to the relation between the mass and the Yukawa coupling
of the bottom quark. As a result of this special choice of parameters,
all terms of the form $\epsilon_b\,\tan\beta$ drop out of the two-loop
part of the form factor.

\subsection{Contributions controlled by the Higgs-sbottom 
coupling}
\label{sec:sbotcontr}

We now turn our attention to the diagrams that involve the
Higgs-sbottom coupling. In our approximation those diagrams contribute
only to the functions $F_b,\, \widetilde F_b$ and $\widetilde G_b$ in
eqs.~(\ref{eq:H1}) and (\ref{eq:H2}).  In analogy with ref.~\cite{DS},
the two-loop parts of the functions can be written as
\bea
F_b^{2\ell} &=& Y_{\sbu} - Y_{\sbd} -\frac{4\,\cdb^2}{\bu-\bd}\, Y_{\cdb^2}
\label{Fderiv}\,,\\
\widetilde F_b^{2\ell} &=& Y_{\sbu} - Y_{\sbd}
+\frac{4\,\sdb^2}{\bu-\bd}\,Y_{ \cdb^2}\,,
\label{Ftderiv} \\
\widetilde G_b^{2\ell} &=& Y_{\sbu} + Y_{\sbd}\,.
\label{Gtderiv}
\eea
Furthermore, the various terms in
eqs.~(\ref{Fderiv})--(\ref{Gtderiv}) can be split in the contributions
coming from diagrams with gluons ($g$, fig.~\ref{fig:diaghsq}a),
with strong, D-term-induced quartic bottom couplings ($4\tilde b$,
fig.~\ref{fig:diaghsq}b), and with gluinos ($\tilde g$, 
fig.~\ref{fig:diaghsq}c),
\be
Y_x = Y_x^{g} + Y_x^{4\tilde b} + Y_x^{\gl}~~~~~~~(x=\sbu,\sbd,\cdb^2)~.
\label{Ycon}
\ee
The first two terms in eq.~(\ref{Ycon}) can be obtained from the
analytic expressions presented for the stop contributions in
eqs.~(27)--(30) of ref.~\cite{DS}, identifying $\frac{\partial
  \gam^a}{\partial x}$ with $Y^a_x$ after making the trivial
replacement $\tilde t \rightarrow \tilde b$.
The gluino contributions, on the other hand, require a dedicated
calculation. Writing
\be
Y_x^{\gl} =  C_F \,Y_x^{(\gl,C_{F})} + C_A \, Y_x^{(\gl,C_{A})}~~~~
                       (x=\sbu,\sbd,\cdb^2)~,
\ee
and assuming that the parameters in ${\cal H}_1^{1\ell}$ and ${\cal
  H}_2^{1\ell}$ are expressed in the $\drbar$ scheme at the
renormalization scale $Q^2$, we find for the functions $Y^{\gl}_x$:
\bea
Y_{\sbu}^{(\gl,C_{F})} &=&  \frac{\sdb}{ 4\,m_b\, \mg}
{\mathcal G}^{1\ell}_{1/2} (\tau_b)
\left( \frac1{1-\x1g} +\frac1{(1-\x1g)^2} \ln \x1g\right)\nonumber \\ 
&& -
\frac1{6\, \mg^2} \left( \frac{1}{1-\x1g}  +
\frac1{(1-\x1g)^2}  \ln \x1g  - \frac1{\x1g^2} 
+\frac1{\x1g^2} \ln \frac{\mg^2}{Q^2} \right)~,
\label{YCFB1}\\
Y_{\sbu}^{(\gl,C_{A})} &=& -\frac{1}{12\, \mg^2}
\left( \frac1{1-\x1g} +\frac1{(1-\x1g)^2} \ln \x1g\right) ~,
\label{YCAB1}   \\
Y_{\cdb^2}^{(\gl,C_{F})} &=& -\frac{\mg}{8\,m_b \sdb}\, 
{\mathcal G}^{1\ell}_{1/2} (\tau_b) \,\left( \frac{\x1g}{1-\x1g} \ln \x1g
-   \frac{x_2}{1-x_2} \ln x_2 \right)~,
\label{YCFc2} \\
Y_{\cdb^2}^{(\gl,C_{A})} &=& 0~,
\label{YCAc2}
\eea
where we retained only terms that induce ${\cal O}(m_b^2/m_h^2)$,
${\cal O}(\mb/M)$ and ${\cal O}(\mz^2/M^2)$ contributions to
${\mathcal H}^{2\ell}_{1,2}$. The expression for $Y_{\sbd}$ can be
obtained from the expression for $Y_{\sbu}$ through the replacements
$x_1 \rightarrow x_2$ and $\sdb \rightarrow-\sdb$.  Comparing
eqs.~(\ref{YCFB1}) and (\ref{YCAB1}) with eq.~(42) of ref.~\cite{DS}
we notice that, contrary to what we stated in section 3.3 of that
paper, even for $\theta_b=0$ the two-loop bottom-sbottom-gluino
contribution to $\widetilde F_b$ and $\widetilde G_b$ {\em cannot} be
obtained by taking the limit $\mt\rightarrow 0$ in the corresponding
top contribution.

\subsection{On-shell renormalization scheme for the sbottom
  parameters}
\label{sec:OS}

We now discuss a suitable OS renormalization scheme for the parameters
that determine the sbottom contribution to ${\cal H}_{1,2}^{1\ell}$.
We recall that, at the one-loop level, the vev $v_1$ and $v_2$, the
$Z$ boson mass, the Weinberg angle and the parameter $\mu$ are not
renormalized by the strong interactions. Therefore, the only
parameters that require a one-loop definition are ($h_b, \, A_b,\,
\sdb, m_{\sbu} ,\, m_{\sbd}$), where by $h_b$ we denote the coupling
constant entering the cubic and quartic sbottom-Higgs interactions,
which at tree level is related to the bottom mass by $\mb =
h_b\,v_1/\,\sqrt{2}$.  Indeed, the factor $\mb$ that multiplies the
function $F_b$ in eqs.~(\ref{eq:H1}) and (\ref{eq:H2}) has to be
interpreted as a bookmark for $h_b$. In fact, only four of those
parameters are independent, because of the relation
\be
\sdb = 
\frac{\sqrt{2} \, h_b  \,( A_b \, v_1 + \mu \, v_2 )}{\bu - \bd }  \, .
\label{s2b}
\ee
In the analysis of the sbottom corrections to the neutral MSSM Higgs
boson masses presented in ref.~\cite{bdsz2} it was pointed out that,
while the sbottom masses can be naturally identified with the pole
masses, an OS definition of $(h_b,A_b,\sdb)$ is less easily singled
out. Proceeding in analogy with the OS renormalization of the stop
sector (see, e.g., ref.~\cite{DS}), we might choose as independent
parameters a conveniently defined bottom mixing angle, $\sdb$, and the
bottom Yukawa coupling $h_b^{pole}$, as defined by the pole bottom
mass $M_b$ via the relation $M_b \equiv h_b^{pole} v_1/
\sqrt{2}$. Then, eq.~(\ref{s2b}) might be used to establish the
one-loop definition of $A_b$ in terms of the pole bottom and sbottom
masses and the sbottom mixing angle. However, for large values of
$\tan\beta$ such definition would produce very large shifts in $A_b$
with respect to its $\drbar$ value\,\footnote{For the generic
  parameter $x$, we define the shift from the $\drbar$ value $\hat{x}$
  as $\delta x \equiv \hat{x} - x$.}, $\delta A_b = {\cal O}(\as \,
\mu^2 \, \tan^2 \beta / \mg)$ \cite{eberl}.  This is related to the
fact that, in the large-$\tb$ limit (i.e., $v_1 \to 0$), $\sdb$
becomes independent of $A_b\,,$ as can be easily seen from
eq.~(\ref{s2b}). To cure the problem, it was suggested in
ref.~\cite{bdsz2} (see also ref.~\cite{heidi}) to take $\sdb$ and
$A_b$ as independent parameters, while considering $h_b$ as a derived
quantity via eq.~(\ref{s2b}).  Suitable renormalization conditions
were then proposed for $\sdb$ and $A_b$.

In the OS analysis of the cross section for Higgs boson production we
want to retain the convenient features of the renormalization
prescription employed in ref.~\cite{bdsz2}. However, that prescription
needs to be expanded: first of all, the renormalization conditions in
ref.~\cite{bdsz2} were defined in the limit $\tb \to \infty$ (i.e.,
$v_1=0$), while in the case at hand we do not impose constraints on
$\tb$. Moreover, while in the calculation of the one-loop corrections
to the Higgs masses the contributions controlled by the bottom-Higgs
Yukawa coupling (which we denote as $h_b^{Y}$ to distinguish it from
$h_b$) are suppressed, the one-loop diagram controlled by $h_b^{Y}$
gives an important contribution to the production cross
section. Therefore, a one-loop definition of $h_b^{Y}$ is
required. Since it does not seem appropriate to define the
bottom-Higgs coupling $h_b^{Y}$ in terms of quantities of the sbottom
sector, as would happen if we imposed on it the same renormalization
condition used for $h_b$, we need to impose different renormalization
conditions on $h_b^Y$ and $h_b$, or, equivalently, on the bottom mass
that enters the one-loop bottom contribution and the one that enters
the one-loop sbottom contribution. In particular, we identify the
former with the pole mass $M_b$ (the resulting shifts in the function
${G}_b^{2\ell}$ are discussed in section
\ref{sec:botcontr}). Concerning the bottom mass in the one-loop
sbottom contribution, we follow ref.~\cite{bdsz2}, extending the
prescription presented in that paper to the case of finite $\tb$.

To obtain definitions for $\delta h_b$ and $\delta A_b$, we consider
two quantities
\be
\widetilde{X}_b  =  {h_b \, v_1 \over \sqrt{2}} \,(  A_b + \mu\,\tb)~,
\;\;\;\;\;~~~~ 
\widetilde{Y}_b  =   {h_b \over \sqrt{2}} \,(\sb \, A_b - \cb \, \mu)~, 
\label{XY}
\ee
that allow for a natural interpretation: $\widetilde{X}_b\,$, at the
classical level, is the off-diagonal term in the sbottom mass matrix,
related to the mixing angle $\sdb$ via eq.~(\ref{s2b});
$\widetilde{Y}_b$ is proportional to the coefficient of the trilinear
interaction $(\widetilde{b}_1 \widetilde{b}_2^* A)$. A definition of
the mixing angle $\theta_b$ like the one proposed in
ref.~\cite{mixing},
\be
\delta \theta_b = \frac12\,
\frac{\widehat{\Pi}_{12}(\bu)+\widehat{\Pi}_{12}(\bd)}{\bu-\bd}~,
\label{dthus}
\ee
together with the identification of the sbottom masses as pole masses,
can be immediately translated, using eq.~(\ref{s2b}), into a
prescription for $\widetilde{X}_b$:
\be
\label{deltaXb}
\delta \widetilde{X}_b = {1 \over 2} c_{2\theta_{b}} \left[ \widehat{\Pi}_{12}
(\bu) + \widehat{\Pi}_{12}(\bd) \right] + \widetilde{X}_b \, 
\frac{\widehat{\Pi}_{11}(\bu) - \widehat{\Pi}_{22}(\bd)} {\bu- \bd} \, .
\ee
In eqs.~(\ref{dthus}) and (\ref{deltaXb}), $\widehat{\Pi}_{ij}(q^2)\:
(i,j=1,2)$ denotes the finite part of the $(i,j)$ self-energy of the
sbottoms.

Recalling that in the $\tb \to \infty$ limit $\widetilde{Y}_b \to h_b
\, A_b / \sqrt{2}$, the extension to the case of finite $\tb$ of the
prescription for $A_b$ introduced in eq.~(15) of ref.~\cite{bdsz2}
reads:
\bea
\label{deltaYb}
\delta\widetilde{Y}_b & = & - {i \over 2}\left[
\Lambda_{12A}(\bu,\bu,0) + \Lambda_{12A}(\bd,\bd,0) \right] \nn\\
&&\nn\\
& & + {1 \over 2}\, \widetilde{Y}_b \, 
\frac{ \widehat{\Pi}_{11}(\bu)+ \widehat{\Pi}_{22}(\bu) 
- \widehat{\Pi}_{11}(\bd) - \widehat{\Pi}_{22}(\bd)} {\bu- \bd} ~
\eea
where $i \Lambda_{12A}(p_1^2, p_2^2, p_A^2)$ denotes the proper vertex
$\tilde{b}_1 \tilde{b}_2^* A$.

Finally, the shifts of the parameters $h_b$ and $A_b$ are related to
those of $\widetilde{X}_b$ and $\widetilde{Y}_b$ by
\bea 
\delta h_b &=& {\sqrt{2} \over {\mu\, v}} \left( \delta\widetilde{X}_b \,
\sb - \delta\widetilde{Y}_b \, v \,\cb \right) ~,
\label{dhb} \\
\delta A_b & = & 
{ 2 \,\over h_b^2\, \mu\, v} \left(
\widetilde{X}_b\, \delta\widetilde{Y}_b 
- \widetilde{Y}_b \, \delta\widetilde{X}_b  \right)~,
\label{deltaab}
\eea
where $v = \sqrt{v_1^2+v_2^2}$. Explicit expressions for $\delta h_b$
and $\delta A_b$, as well as for $\delta\sdb$ and $\delta m_{\tilde
  b_i}^2$, can be found in appendix B.

If the one-loop sbottom contribution to ${\cal H}_1^{1\ell}$ and
${\cal H}_2^{1\ell}$ is evaluated in terms of OS quantities, the
two-loop functions in eqs.~(\ref{Fderiv})--(\ref{Gtderiv}) must be
replaced by

\bea 
F_b^{2\ell} & \longrightarrow & F_b^{2\ell} ~+~
\frac{\pi}{6\,\alpha_s}\,\left[
\frac{\delta\bu}{\buq}-\frac{\delta\bd}{\bdq}-\left(
\frac{\delta h_b}{h_b}+\frac{\delta\sdb}{\sdt}\right)
\,\left(\frac{1}{\bu}-\frac{1}{\bd}\right)\right]~,\label{shiftF}\\
&&\nn\\
\label{shiftFt}
\widetilde F_b^{2\ell} & \longrightarrow & \widetilde F_b^{2\ell} ~+~
\frac{\pi}{6\,\alpha_s}\,\left[
\frac{\delta\bu}{\buq}-\frac{\delta\bd}{\bdq}
- \frac{\delta\cdb}{\cdb}\,\left(\frac{1}{\bu}-\frac{1}{\bd}\right)\right]~,\\
&&\nn\\
\label{shiftGt}
\widetilde G_b^{2\ell} & \longrightarrow & \widetilde G_b^{2\ell} ~+~
\frac{\pi}{6\,\alpha_s}\,\left[
\frac{\delta\bu}{\buq}+\frac{\delta\bd}{\bdq}\right]~.
\eea
In addition, the two-loop form factor ${\cal H}_1^{2\ell}$ receives a
contribution originating from the shift in $A_b$:
\be
{\cal H}_1^{2\ell} \longrightarrow {\cal H}_1^{2\ell} ~-~
\frac{\mb\,\sdb}{c_\beta}\,
\frac{\pi}{6\,\alpha_s}
\,\left(\frac{1}{\bu}-\frac{1}{\bd}\right)\,\delta A_b~.\label{shiftA}
\ee


\section{A numerical example}
\label{sec:num}

We will now illustrate the effect of the two-loop bottom/sbottom
contributions to the form factors for the production of a Higgs boson
in a representative region of the MSSM parameter space.

The SM parameters entering our calculation are the $Z$ boson mass $\mz
= 91.1876$ GeV, the Fermi parameter $G_F = 1.16637\times10^{-5}$
GeV$^{-2}$, the sine of the Weinberg angle $s_{\theta_W}^2 = 0.223$
and the strong coupling constant $\alpha_s(\mz) = 0.118$
\cite{PDG}. For the pole masses of the top and bottom quarks we take
$M_t = 173.1$ GeV \cite{topmass} and $M_b = 4.49$ GeV, the latter
corresponding to the SM running mass (in the $\msbar$ scheme)
$\overline \mb(\mb) = 4.16$ GeV \cite{botmass}.  The tree-level mass
matrix for the CP-even Higgs bosons can be expressed in terms of the
physical pseudoscalar mass $m_A$ and the $\drbar$-renormalized
parameter $\tan\beta$, in addition to $\mz$. In the calculation of the
physical Higgs boson masses and of the mixing angle $\alpha$ we
include the one-loop ${\cal O}(\alpha_t +\alpha_b)$ and two-loop
${\cal O}(\alpha_t\alpha_s+\alpha_b\alpha_s)$ corrections as in
refs.~\cite{bdsz2,DSZ}.

When computing the two-loop corrections to both mass matrix and
production form factors for the CP-even Higgs bosons, the parameters
that determine the stop and sbottom masses and mixing angle and are
subject to ${\cal O}(\alpha_s)$ corrections require a one-loop
specification. For the stop sector we adopt the OS scheme described
e.g.~in ref.~\cite{DS}. In particular, we take as input the pole
top mass $M_t$ and the soft SUSY-breaking parameters
$(m_{Q,\tilde{t}}\,, m_U, A_t)$ that can be derived by rotating the
diagonal matrix of the OS stop masses by the angle $\theta_t$, defined
as in eq.~(37) of ref.~\cite{DS}. Concerning the corresponding
parameters of sbottom sector $(h_b, m_{Q,\tilde{b}}\,, m_D, A_b)$
additional care is required, because of our non-trivial definition of
$h_b$ and of the fact that, at ${\cal O}(\as)$, the parameter
$m_{Q,\tilde{b}}$ entering the sbottom mass matrix differs from the
corresponding stop parameter $m_{Q,\tilde{t}}$ by a finite shift
\cite{eberl}.  We start by computing the renormalized Higgs-sbottom
coupling as given by $h_b = \hat h_b - \delta h_b$, where $\hat h_b$
is the $\drbar$-renormalized running coupling that can be trivially
extracted from $\widehat \mb$  computed via
eq.~(\ref{mbrun}), and $\delta h_b$ is defined in
eq.~(\ref{dhb}). Then we compute $m_{Q,\tilde{b}}$ following the
prescription of \cite{eberl}.  Finally, we use the parameters $h_b$
and $m_{Q,\tilde{b}}$ to compute the actual values of the OS sbottom
masses and mixing angle.

\begin{figure}[t]
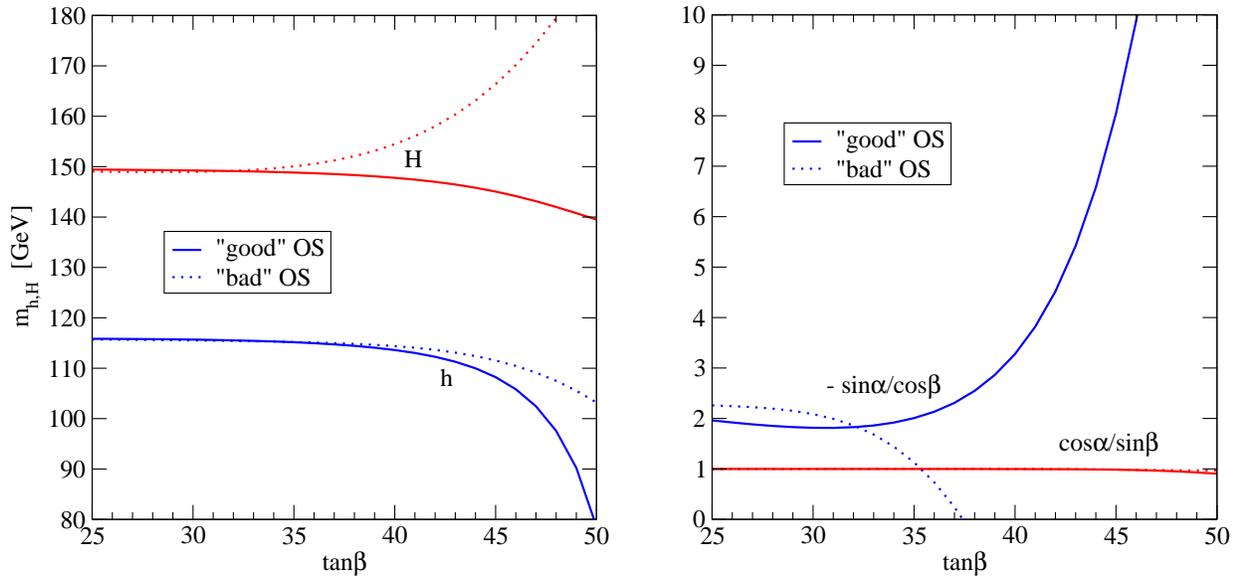

\begin{center}
\mbox{
\epsfig{figure=mhtb.eps,width=8cm}~~~~~~
\epsfig{figure=abtb.eps,width=7.35cm}
}
\caption{CP-even Higgs boson masses (left plot) and effective
  couplings of $h$ to top and bottom quarks (right plot) as a function
  of $\tan\beta$, for $m_A = 150$ GeV and all SUSY mass parameters
  equal to $M=500$ GeV. For the meaning of the solid and dotted lines
  see the text.}
\label{fig:massmix}
\end{center}
\end{figure}

To set the stage for further discussion, we show in
fig.~\ref{fig:massmix} the masses and mixing angle of the CP-even
Higgs bosons as a function of $\tan\beta$. All the relevant
SUSY-breaking parameters, as well as the supersymmetric mass parameter
$\mu$, are set to a common value $M$ = 500 GeV, and the physical
pseudoscalar mass $m_A$ is set to 150 GeV.  The left panel of
fig.~\ref{fig:massmix} shows the masses of the two Higgs bosons $h$
and $H$ in the range $25<\tan\beta<50$, while the right panel shows
the combinations $-\sin\alpha/\cos\beta$ and $\cos\alpha/\sin\beta$,
which determine the strength of the coupling of $h$ to the bottom and
top quarks, respectively, relative to the corresponding SM
couplings. For each set of curves, the solid line represents the
result obtained in the OS renormalization scheme of ref.~\cite{bdsz2},
described in section \ref{sec:OS}. For comparison, we also show as a
dotted line the result that would be obtained if the sbottom
parameters $h_b$ and $A_b$ were renormalized in the same way as the
corresponding stop parameters. The left plot shows the well-known fact
that, at large $\tan\beta$, the radiative corrections from sbottom
loops tend to reduce $m_h$. The right plot shows that, for the chosen
values of $m_A$ and $\tan\beta$, the coupling of $h$ to the bottom
quark is still substantially enhanced with respect to its SM
value. This has to be contrasted with the couplings of $h$ to the top
quark and to the gauge bosons (not shown), which are already very
close to the SM values they tend to in the ``decoupling'' limit $m_A
\gg \mz$.

The comparison between the dotted and solid lines in
fig.~\ref{fig:massmix} shows that, if we had adopted for the sbottom
parameters $h_b$ and $A_b$ the renormalization scheme used for the
stop parameters, the results for $m_H$ and for $-\sin\alpha/\cos\beta$
would differ wildly from the ones obtained with the renormalization
scheme discussed in section \ref{sec:OS} (conversely, we checked that
the results obtained in the $\drbar$ scheme would be in good
qualitative agreement with the solid lines). The discrepancy is due to
the fact that in the ``bad'' OS scheme the (1,1) and (1,2) entries of
the CP-even Higgs mass matrix are subject to very large two-loop
corrections scaling like $M^2\tan\beta^2$, induced by the contribution
of the counterterm $\delta A_b$. It is interesting to note that, since
the contribution of $\delta A_b$ to the form factor ${\cal
  H}_1^{2\ell}$ in eq.~(\ref{shiftA}) is suppressed by a factor $\mb$,
its impact on the Higgs boson production cross section in the ``bad'' OS
scheme is not as extreme as the impact on the Higgs mass. However, we
stress that a consistent determination of the properties of the Higgs
bosons requires that the same definition of input parameters be used
in the calculations of mass and production cross section. Since the
naive choice of using the same OS renormalization scheme for the stop
and sbottom sectors is not viable in the calculation of the Higgs
masses, it should not be applied to the calculation of the cross
section either.

\begin{figure}[t]
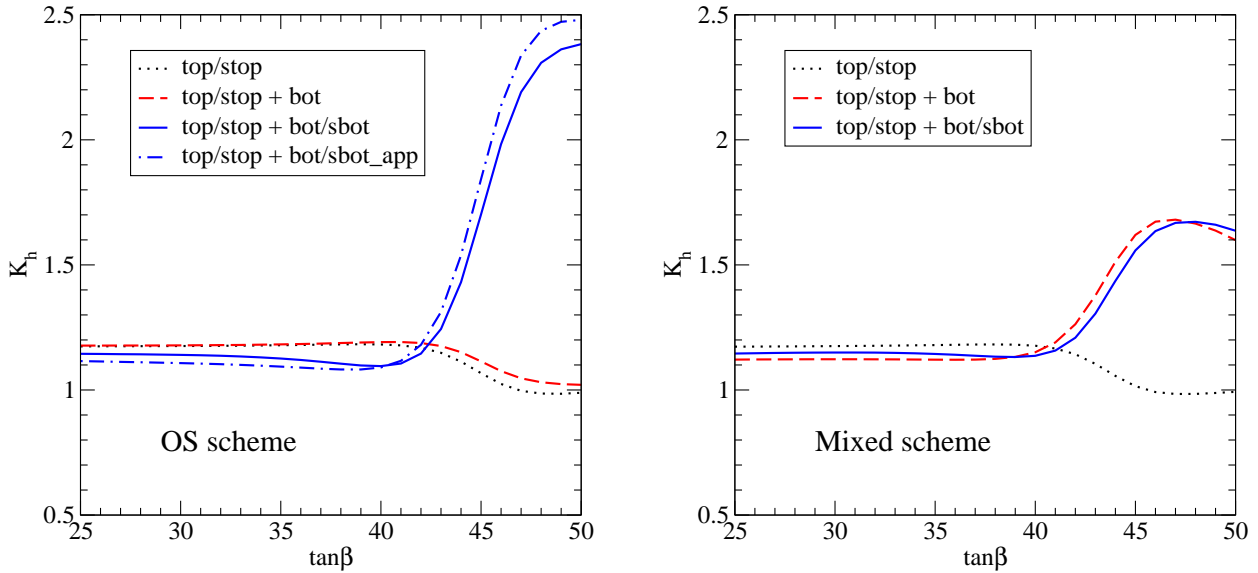

\begin{center}
\mbox{
\epsfig{figure=KfacOS.eps,width=7.8cm}~~~~~~
\epsfig{figure=KfacOR.eps,width=7.8cm}
}
\caption{$K$ factor for the production of a light Higgs boson $h$ as a
  function of $\tan\beta$, for $m_A = 150$ GeV and all SUSY mass
  parameters equal to $M=500$ GeV. For the meaning of the different
  lines see the text.}
\label{fig:kfactorsh}
\end{center}
\end{figure}

We are now ready to discuss the effect of the two-loop bottom/sbottom
contributions to the form factor for Higgs-boson production. To this
purpose, we define a factor $K_h$ that contains the ratio of two-loop to
one-loop form factors appearing in eq.~(\ref{ggg}):
\be
\label{kfac}
K_h ~=~ 1\,+ \,2\,\frac{\alpha_s}{\pi}\,{\rm Re}
\left(\frac{-\sin\alpha \,{\mathcal H}^{2\ell}_1 
+\cos\alpha \,{\mathcal H}^{2\ell}_2}{
-\sin\alpha \,{\mathcal H}^{1\ell}_1 
+ \cos\alpha \,{\mathcal H}^{1\ell}_2 }\right)~.
\ee

In the left panel of fig.~\ref{fig:kfactorsh} we plot $K_h$ as a
function of $\tan\beta$, with the same choice of SUSY parameters as in
fig.~\ref{fig:massmix}, in the OS renormalization scheme described in
section \ref{sec:OS}. The one-loop form factors in the denominator of
the term between parentheses in eq.~(\ref{kfac}) contain both the
top/stop and bottom/sbottom contributions, computed under the
approximations of eqs.~(\ref{Glimit}) and (\ref{GLimit}). The lines in
the plot correspond to different computations of the two-loop form
factors in the numerator: the dotted line includes only the
contributions of the top/stop sector, as computed in ref.~\cite{DS};
the dashed line includes also the contribution of two-loop diagrams
with bottom quarks and gluons; the solid line includes the full
two-loop contribution of the bottom/sbottom sector as computed in
section \ref{sec:2loopres}; finally, the dot-dashed line is obtained
by approximating the bottom/sbottom contribution (with the exception
of the bottom-gluon diagrams) with just the terms enhanced by
$\mg/\mb$ in eq.~(\ref{YCFG}). From the comparison between the dotted
and dashed lines it can be seen that, in the OS renormalization
scheme, the contribution to ${\cal H}_1^{2\ell}$ of the two-loop
diagrams with bottom quarks and gluons is very small. This is due to a
partial cancellation between the terms $C_F\,{\cal
  F}^{(2\ell,a)}_{1/2}$ and $C_A\,{\cal G}^{(2\ell,C_A)}_{1/2}$
entering the function $G_b^{2\ell}$, and to the fact that, in this
scheme, the term ${\cal F}^{(2\ell,b)}_{1/2}$ is not enhanced by the
potentially large logarithm of the ratio between the bottom mass and
the renormalization scale, as can be seen by comparing
eqs.~(\ref{gdl}) and (\ref{GbOS}). The solid line shows that the
effect of the diagrams involving sbottoms can be very sizable at large
$\tan\beta$, more than doubling $K_h$.  Indeed, for large $\tan\beta$
the coupling of the light Higgs boson to the (s)bottom is considerably
enhanced with respect to its SM value, as can be seen in the right
panel of fig.~\ref{fig:massmix}.  However, the proximity between the
solid and dot-dashed lines shows that this sizable effect is almost
entirely due to the terms enhanced by $\mg/\mb$ in the contribution of
the two-loop bottom-sbottom-gluino diagrams in which the light Higgs
boson couples to the bottom quark.

As discussed in section \ref{sec:botcontr}, the terms enhanced by
$\mg/\mb$ in the OS result can be canceled out if the Higgs-bottom
Yukawa coupling in the one-loop part of the result is identified with
the $\drbar$-renormalized MSSM bottom mass $\widehat m_b$ instead of
the physical mass $M_b$. To this effect, the factor $m_b^2$
multiplying the function $G_b$ in eq.~(\ref{eq:H1}) must be expressed
as $\widehat m_b\,M_b$, and the two-loop part of $G_b$ must be shifted
as in eq.~(\ref{shift}). In the right panel of
fig.~\ref{fig:kfactorsh} we present the result of this manipulation,
with $\widehat m_b$ evaluated at the scale $Q=m_h$.  The input
parameters and the meaning of the different lines are the same as for
the plot in the left panel.  The proximity between the dashed and
solid lines shows that the contribution of the two-loop diagrams
involving sbottoms is rather small in this renormalization scheme, at
least for our choice of input parameters. However, $K_h$ still shows a
sizable increase at large $\tan\beta$. This is due to the fact that
the shift in eq.~(\ref{shift}) brings back a large logarithm, $\ln
(\mb^2/m_h^2)$, in the contribution of the two-loop diagrams with
bottom and gluon (this logarithm compensates the scale dependence of
the running mass $\widehat m_b$).

\begin{figure}[t]
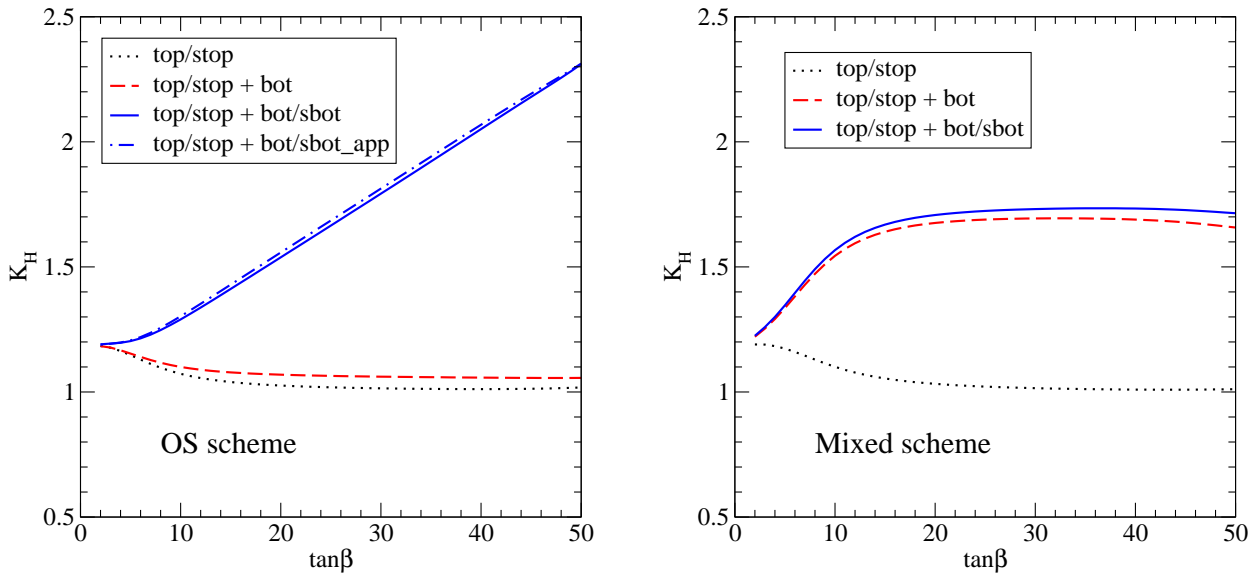

\begin{center}
\mbox{
\epsfig{figure=KfacOSH.eps,width=7.8cm}~~~~~~
\epsfig{figure=KfacORH.eps,width=7.8cm}
}
\caption{Same as figure \ref{fig:kfactorsh} for the heavy Higgs boson $H$.}
\label{fig:kfactorshh}
\end{center}
\end{figure}

To conclude this section, we show in fig.~\ref{fig:kfactorshh} the
factor $K_H$ for the production of the heavy CP-even Higgs boson, in
the range $2 < \tan\beta < 50$. The definition of $K_H$ can be
obtained from the one of $K_h$ in eq.~(\ref{kfac}) via the
replacements $\sin\alpha \rightarrow -\cos\alpha,~\cos\alpha
\rightarrow\sin\alpha$. The input parameters are chosen exactly as in
figs.~\ref{fig:massmix} and \ref{fig:kfactorsh}, and the meaning of
the lines in the left and right panels is the same as in
fig.~\ref{fig:kfactorsh}. Since in this example the mass of the heavy
Higgs boson is of the order of 150 GeV (see fig.~\ref{fig:massmix}),
i.e.~well below any threshold for heavy-particle production, we expect
the approximation of vanishing Higgs mass to hold reasonably well even
for $H$. From fig.~\ref{fig:kfactorshh} it appears that the balance of
the various contributions to $K_H$ in the two different
renormalization schemes is qualitatively similar to the one for $K_h$
shown in fig.~\ref{fig:kfactorsh}: in the OS scheme 
 the factor $K_H$ receives a sizable contribution from the
sbottom diagrams, largely dominated by the terms enhanced by $\mg/\mb$
in the diagrams controlled by the Higgs-bottom coupling; in the
``mixed'' scheme, on the other hand, the sbottom contribution is
rather small, but there is a sizable contribution from the diagrams
with bottom and gluon.

This said, the factor $K_H$ shows a peculiar dependence on
$\tan\beta$: for sufficiently large values of $\tan\beta$, it grows
linearly in the OS scheme, while it reaches a plateau in the mixed
scheme. This can be easily understood by recalling that, for
moderate-to-large $\tan\beta$ and for our choice of $m_A$, the Yukawa
coupling of the heavy MSSM Higgs to bottom quarks is enhanced by
$\tan\beta$ with respect to the SM value, while the coupling to top
quarks is suppressed by $\tan\beta$.  Consequently, both the one-loop
and the two-loop form factors in $K_H$ are dominated by the
contribution of the diagrams controlled by the Higgs-bottom coupling,
with the result that the coupling itself cancels out in the
ratio. However, the dominant contribution from the
bottom-sbottom-gluino diagrams in the OS scheme contains an additional
$\tan\beta$-enhancement hidden in the product $\sdb\,\mg/\mb$ (see the
discussion at the end of section \ref{sec:botcontr}), which explains
the linear rise of $K_H$. On the other hand, the dominant contribution
of the bottom-gluon diagrams in the mixed scheme possesses no further
$\tan\beta$-enhancement, which explains the plateau.


\section{Conclusions and discussion}
\label{sec:concl}

In this paper we presented analytic results for the NLO
bottom-sbottom-gluino contribution to the cross section for
Higgs boson production in gluon fusion, obtained using an asymptotic
expansion in the large supersymmetric masses. This approximation is
fully valid for the light-Higgs case, while for the heavy Higgs it
covers the mass region where $m_H$ is below all the heavy-particle
thresholds. Together with the previously known results for the NLO
corrections in the MSSM, our expressions can be easily implemented in
computer codes that aim to provide an accurate and efficient
evaluation of the cross section for Higgs boson production in the
MSSM.

In our analysis we paid special attention to the consistency between
the calculations of the masses and the production cross sections of
the MSSM Higgs bosons, i.e.~to the fact that the same input
parameters, defined at the one-loop level, should be used in both
calculations.  The OS definition of the parameters of the bottom
sector is delicate, as discussed for the case of the Higgs masses in
ref.~\cite{bdsz2}. The choice of treating the top and the bottom
sectors on the same footing suffers from the fact that large two-loop
corrections proportional to $\tan^2\beta$ are generated in the
contributions controlled by the Higgs-sbottom couplings, affecting
both the calculation of the Higgs masses and that of the production
cross sections.  To avoid such large two-loop effects, a convenient OS
renormalization prescription was proposed in ref.~\cite{bdsz2} for the
calculation of the Higgs masses. In the present paper we have extended
that prescription to cover also the calculation of the production
cross sections.

Our analysis of the NLO bottom contribution to the gluon-fusion
production cross section shows that, with our choice of OS
renormalization conditions, the bulk of the corrections comes from the
two-loop diagrams involving the Higgs-bottom Yukawa coupling, while
the diagrams controlled by the Higgs-sbottom coupling play a secondary
role. The contribution controlled by the Higgs-bottom Yukawa coupling
can be further divided in two parts: diagrams with only bottom and
gluons and diagrams involving bottom, sbottom and gluino. By far, the
most important pieces of the latter diagrams are the terms of ${\cal
  O}( \mb \mg/m^2_{h,H})$, i.e.~the ones in which the helicity flip on
the fermion line is achieved via a gluino mass insertion instead of a
bottom mass insertion.

It is natural to wonder if it is possible to absorb most of the NLO
bottom contribution into the LO term with a suitable choice of the
input parameters. In such a situation the factor $K_{h,H}$, as defined
in eq.~(\ref{kfac}), would be basically sensitive to the top/stop
contribution only.  The contribution of the two-loop bottom-gluon
diagrams can be made small if the one-loop bottom diagrams are
expressed in terms of the pole bottom mass $M_b$, but in this case the
${\cal O}( \mb \mg/m^2_{h,H})$ terms give a sizable contribution.  On
the other hand, if the Higgs-bottom Yukawa coupling in the one-loop
result is expressed in terms of the running bottom mass
$\widehat{m}_b$, and the bottom mass in terms of $M_b$, the ${\cal O}(
\mb \mg/m^2_{h,H})$ terms in the two-loop contribution cancel out, but
the bottom-gluon diagrams give a relevant contribution because of the
presence of large logarithms of the ratio between $\mb$ and the
renormalization scale. However, the explicit knowledge of the NLO
bottom contribution allows us to devise a simple recipe to absorb the
bulk of the NLO contribution into the LO term.  It amounts to writing
the LO bottom contribution entirely in terms of the pole bottom mass
$M_b$, then rescaling it by a factor $1/(1 + \epsilon_b \tan
\beta)$. Once this manipulation is implemented, we expect the
remaining NLO bottom/sbottom contributions to be quite small -- at
least in large regions of the parameter space -- in which case they
can be neglected in the evaluation of the form factors ${\cal
  H}_{1,2}^{2\ell}$ without introducing large errors\footnote{A
  somewhat similar procedure was suggested, without a detailed
  discussion, in ref.~\cite{spiraDb}.}.  We stress that the validity
of this simple recipe is strictly linked to the absence of spuriously
large corrections to the Higgs-sbottom coupling. This is realized with
our choice of OS renormalization conditions for the sbottom sector
(and also in the $\drbar$ scheme) but it is not guaranteed with other
renormalization conditions.

Finally, the results derived in this paper for the production cross
section can be straightforwardly applied to the NLO computation of the
gluonic and photonic decay widths of the CP-even Higgs boson in the
MSSM, as described in section 5 of ref.~\cite{DS}.


\section*{Acknowledgments}
This work was supported in part by an EU Marie-Curie Research Training
Network under contract MRTN-CT-2006-035505 (HEPTOOLS) and by ANR under
contract BLAN07-2\_194882.


\section*{Appendix A: NLO contributions from real parton emission}
\label{appA}
\begin{appendletterA}

In this appendix we specialize to the MSSM case the general exact
results of ref.~\cite{BDV} for the functions ${\cal R}_{gg},\, {\cal
  R}_{q \bar q},\, {\cal R}_{q g}$. We aim at expressions that, on one
hand, are sufficiently accurate, while on the other hand allow for a
fast numerical evaluation. Thus we report expressions in which the
contributions of the top quark and of the squarks are evaluated in the
limit of neglecting the Higgs mass, while the contribution of the
bottom quark is kept exact.

The function ${\cal R}_{gg}$ can be written as
\be
{\cal R}_{gg} = \frac1{z(1-z)}\int_0^1 \frac{d v}{v (1-v)} \left\{
 \frac{8\,z^4 \left| {\cal A}_{gg}(\hat{s},\hat{t},\hat{u})\right|^2 }{  
\left| -\sin\alpha \,{\mathcal H}^{1\ell}_1 +
\cos\alpha \,{\mathcal H}^{1\ell}_2 \right|^2 }  - (1-z+z^2)^2 \right\} ,
\label{eqA1}
\ee
where $\hat{t} = -\hat{s} (1-z) (1-v),\,\hat{u} = -\hat{s} (1-z) v$,
with
\be
\left| {\cal A}_{gg}(s,t,u) \right|^2  =
 |A_2 (s,t,u)|^2 + |A_2 (u,s,t)|^2 + |A_2 (t,u,s)|^2 +
      |A_4 (s,t,u)|^2 .
\label{Agg}
\ee
Furthermore, the functions $A_2$ and $A_4$ can be cast in the following form:
\bea
A_2 (s,t,u) & = &  -\sin\alpha \,{\mathcal R}^{A_{2}}_1 (s,t,u) +
\cos\alpha \,{\mathcal R}^{A_{2}}_2 (s,t,u)~,\\
A_4 (s,t,u) & = &  -\sin\alpha \,{\mathcal R}^{A_{4}}_1 (s,t,u) +
\cos\alpha \,{\mathcal R}^{A_{4}}_2 (s,t,u) ~,
\eea
with 
\bea
{\mathcal R}^{A_{2}}_1 (s,t,u) & = & 
\frac{s^2}{4\, (s+t+u)^2}\,  {\mathcal H}^{1\ell}_1 \nn \\
&+&  \lambda_b \, \left\{ \frac{\tau_b^2}{16} 
\left[ b_{1/2} \left( \frac{s}{m_b^2},\frac{t}{m_b^2},\frac{u}{m_b^2} \right) + 
b_{1/2} \left( \frac{s}{m_b^2},\frac{u}{m_b^2}, \frac{t}{m_b^2} \right) 
\right] \right.
\left. -\frac{s^2}{4\, (s+t+u)^2}\,{\mathcal G}^{1\ell}_{1/2} (\tau_b) 
\right\}~,\nn\\ \label{eqA5} \\
{\mathcal R}^{A_{2}}_2 (s,t,u) & = & 
\frac{s^2}{4\, (s+t+u)^2}\,  {\mathcal H}^{1\ell}_2 ~, \label{eqA6}\\
\nn\\
{\mathcal R}^{A_{4}}_1 (s,t,u) & = & \frac14 \,  {\mathcal H}^{1\ell}_1   \nn \\
& +& \lambda_b \, \left\{  \frac{\tau_b^2}{16} 
\left[ c_{1/2} \left( \frac{s}{m_b^2},\frac{t}{m_b^2},\frac{u}{m_b^2} \right) + 
c_{1/2} \left( \frac{t}{m_b^2},\frac{u}{m_b^2}, \frac{s}{m_b^2} \right) +
c_{1/2} \left( \frac{u}{m_b^2},\frac{s}{m_b^2}, \frac{t}{m_b^2} \right) 
\right] \right. \nn \\
&& ~~~~~~~  - \left. \frac14 {\mathcal G}^{1\ell}_{1/2} (\tau_b) \right\} \, , 
\label{eqA7} \\
{\mathcal R}^{A_{4}}_2 (s,t,u) & = & \frac14 \, {\mathcal H}^{1\ell}_2 \, , 
\label{eqA8}
\eea
where the functions $b_{1/2}(s,t,u)$ and $c_{1/2}(s,t,u)$ are defined
in eqs.~(2.22) and (2.24) of ref.~\cite{BDV}, respectively, and it is
understood that the top and squark contributions to ${\mathcal
  H}^{1\ell}_{1,2}$ are evaluated in the limit of neglecting the Higgs
mass.  In several cases the terms proportional to $\lambda_b$ in
eqs.~(\ref{eqA5}) and (\ref{eqA7}) are numerically very small and can
be neglected.  In such a situation the integration in eq.~(\ref{eqA1})
can be performed analytically, resulting in ${\cal R}_{gg} = -11
(1-z)^3/(6z)$.

The $ q \bar q \to H g $ annihilation channel can be written as
\be
{\cal R}_{q \bar q} = \frac{128}{27} 
 \frac{z\,(1-z)\, \left| {\cal A}_{q \bar q}(\hat{s},\hat{t},\hat{u})\right|^2}
{\left|  -\sin\alpha \,{\mathcal H}^{1\ell}_1 +
\cos\alpha \,{\mathcal H}^{1\ell}_2 \right|^2 }  \, ,
\ee
with
\be
{\cal A}_{q \bar q} ( s,t,u) =   
-\sin\alpha \,{\mathcal R}^{A_{q \bar q}}_1 (s,t,u)
+ \cos\alpha \,{\mathcal R}^{A_{q \bar q}}_2 (s,t,u)~.
\label{eqA10}
\ee
where
\bea
{\mathcal R}^{A_{q \bar q}}_1 (s,t,u) \!& = &\! -\frac{t+u}{2\, (s+t+u)}\,
 {\mathcal H}^{1\ell}_1 
  + \lambda_b \,\left[ \frac{\tau_b}{4} 
d_{1/2} \left( \frac{s}{m_b^2},\frac{t}{m_b^2},\frac{u}{m_b^2} \right)+
\frac{t+u}{2\, (s+t+u)}\, {\mathcal G}^{1\ell}_{1/2} (\tau_b) \right]  ,~~~
\label{eq:Gqq1}\\
{\mathcal R}^{A_{q \bar q}}_2 (s,t,u) \!& = &\! -\frac{t+u}{2\, (s+t+u)}\,
{\mathcal H}^{1\ell}_2 ~.
\label{eq:Gqq2}
\eea
The function $d_{1/2}(s,t,u)$ is defined in eq.~(2.31) of
ref.~\cite{BDV}.

Finally, we consider the quark-gluon scattering channel, $q g \to q H
$. The relevant function ${\cal R}_{qg}$ can be written as
\bea
{\cal R}_{qg} \! \! & = & \! \! C_F \! \int_0^1 \! \frac{d v}{(1-v)} \left\{
\frac{ 1 + (1-z)^2 v^2}{[1-(1-z) v]^2} 
\frac{2 \,z \left| {\cal A}_{qg}(\hat{s},\hat{t},\hat{u})\right|^2 }{  
\left|  -\sin\alpha \,{\mathcal H}^{1\ell}_1 +
\cos\alpha \,{\mathcal H}^{1\ell}_2  \right|^2 }  
   - \frac{1+(1 \! - \! z)^2}{2 z} \right\} ~+~ \frac12\, C_F\, z ~,\nn\\
\label{eqA13}
\eea
where
\be
{\cal A}_{qg}(\hat{s},\hat{t},\hat{u}) = 
{\cal A}_{q\bar q}(\hat{t},\hat{s},\hat{u}) ~.
\ee
As in the case of ${\cal R}_{gg}$, when the terms proportional to
$\lambda_b$ in eq.~(\ref{eq:Gqq1}) can be neglected the integration in
eq.~(\ref{eqA13}) can be performed analytically, giving ${\cal R}_{qg}
= 2 z/3 - (1-z)^2/z$ and ${\cal R}_{q \bar q} = 32 (1-z)^3/(27 z)$.

\end{appendletterA}


\section*{Appendix B: Renormalization scheme shifts in the sbottom sector}
\begin{appendletterB}

  In this appendix we present explicit expressions for the shifts from
  the $\drbar$ to the OS scheme of the parameters in the sbottom
  sector that require a one-loop definition.  Denoting, generically, a
  quantity in the $\drbar$ scheme as $x^{\smalldrbar}$, and the same
  quantity in the OS scheme as $x^{\scriptscriptstyle OS}$, we can
  write the one-loop relation as $x^{\smalldrbar} =
  x^{\scriptscriptstyle OS} + \delta x$. Retaining only terms that do
  not induce contributions suppressed by $\mb^2/M^2$, we find:
\bea
\frac{\delta \bu}\bu & = &  \frac{\alpha_s\,C_F}{4\pi} \,
 \left\{ 3\, \ln \frac{\bu}{Q^2} - 3 -
 \cdb^2 \left( \ln \frac{\bu}{Q^2} -1 \right) 
-\sdb^2 \frac{\bd}{\bu} \left( \ln \frac{\bd}{Q^2} -1 \right)
 \right. \nn \\
&&\nn\\
& & \left. ~~~~~~~~~-6 \frac{\mg^2}{\bu} 
- 2 \left( 1 - 2  \frac{\mg^2}{\bu} \right) \ln \frac{\mg^2}{Q^2}
- 2 \left( 1-\frac{\mg^2}{\bu} \right)^2
\ln \left| 1-\frac{\bu}{\mg^2} \right| \, \right\} \, ,
\label{dmb1}\\
&&\nn\\
\frac{\delta \sdb}\sdb & = & \frac{\alpha_s\,C_F}{4\pi}\, \left\{
-2 \, \cdb^2 +\frac{2\, \cdb^2}{\bu-\bd} \left( \bu   \ln \frac{\bu}{Q^2} 
- \bd  \ln \frac{\bd}{Q^2}  \right)  \right\}  \, ,
\label{thetab} \\
&&\nn\\
\frac{\delta h_b}{h_b} &=& \frac{\alpha_s\,C_F}{4\pi}\,\left\{
-4 + 2\, \ln\frac{\mg^2}{Q^2} 
\phantom{\left(\left(1-\frac{\mg^2}{\bu}\right)^2
\ln \left|1 - \frac{\bu}{\mg^2}\right|\right)}
\right. \nn\\
&&  
\left.
~~~~~~~~~+ \left[ \frac{2\,\bu}{\bu-\bd}\,
\left( 2\,\ln\frac{\bu}{\mg^2} -\left(1-\frac{\mg^2}{\bu}\right)^2
\ln \left|1 - \frac{\bu}{\mg^2}\right|\right)
+\left( 1 \leftrightarrow 2 \right) \right] \right\} \;,
\label{dHb}\\
&&\nn\\
\delta A_b  &=&
\frac{\alpha_s\,C_F}{2\pi} \, \mg \left\{ 
4- 2\, \ln \frac{\mg^2}{Q^2} -\left[ \left( 
1 - \frac{\mg^2}{\bu} \right) \, 
\ln \left|1 - \frac{\bu}{\mg^2} \right|
+ \left( 1 \leftrightarrow 2 \right) \right]
\right\} \, ,
\label{dYb}
\eea
where the notation $(1 \leftrightarrow 2)$ in eqs.~(\ref{dHb}) and
(\ref{dYb}) means a term that is obtained from the previous ones
inside the square bracket with the exchange $\bu \leftrightarrow
\bd$. The shift $\delta \bd$ is obtained from eq.~(\ref{dmb1}) via the
interchange $\bu \leftrightarrow \bd$.  Finally we note that the
expressions for $\delta h_b$ and $\delta A_b$ in eq.~(\ref{dHb}) and
eq.~(\ref{dYb}), respectively, which are valid for generic values of
$\tb$, coincide with the corresponding expressions in
ref.~\cite{bdsz2}, which were derived in the limit $\tb \to \infty$.

\end{appendletterB}


\end{document}